\documentclass[fleqn,usenatbib]{mnras}
\usepackage{graphicx,amsmath,amssymb,url,xspace}
\usepackage{times}
\usepackage{bm}
\usepackage[usenames,dvipsnames]{color}
\usepackage{xfrac}

\newcommand{\Fig}[1]{Figure~\ref{#1}}
\newcommand{\Eq}[1]{Equation~(\ref{#1})}
\newcommand{\Eqs}[2]{Equations~(\ref{#1}) and (\ref{#2})}
\newcommand{\EQ}{\begin{equation}}
\newcommand{\EN}{\end{equation}}
\newcommand{\vv}{\mbox{\boldmath $v$} {}}
\newcommand{\ww}{\mbox{\boldmath $w$} {}}
\newcommand{\GG}{\mbox{\boldmath $g$} {}}
\newcommand{\FF}{\mbox{\boldmath $F$} {}}

\newcommand{\VV}{\mbox{\boldmath $V$} {}}

\newcommand{\DDD}{{\cal D} {}}

\newcommand{\uu}{\mbox{\boldmath $u$} {}}

\newcommand{\Sec}[1]{Section~\ref{#1}}

\newcommand{\ee}{{\bm{\hat{e}}}}

\newcommand{\ez}{{\bm{\hat{e}}_z}}

\newcommand{\nab}{\mbox{\boldmath $\nabla$} {}}

\newcommand{\ov}[1]{\overline{#1}}

\newcommand{\St}{\text{St}}
\newcommand{\Ma}{\text{Ma}}
\newcommand{\TTD}{\text{TTD}}
\newcommand{\alphaSS}{\alpha_{\text{SS}}}

\date{Accepted XXX. Received YYY; in original form ZZZ}

\pubyear{2015}

\title[Turbulent Thermal Diffusion]{Turbulent Thermal Diffusion: A Way to  Concentrate Dust in Protoplanetary Discs}

\author[Alexander Hubbard]{Alexander Hubbard,$^1$\thanks{E-mail: ahubbard@amnh.org} \\
$^{1}$Department of Astrophysics, American Museum of Natural History, New York, NY 10024-5192, USA}

\begin{document}
\label{firstpage}
\pagerange{\pageref{firstpage}--\pageref{lastpage}}
\maketitle

\begin{abstract}
Turbulence acting on mixes of gas and particles generally evenly diffuses the latter through the former. However,
in the presence of background gas temperature gradients
a phenomenon known as turbulent thermal diffusion appears as a particle drift velocity (rather than a diffusive term).
This process moves particles from hot regions to cold ones.
We rederive turbulent thermal diffusion using astrophysical language and demonstrate that it could
play a major role in protoplanetary discs 
by concentrating particles by factors of tens.
Such a concentration would set the stage for collective behavior such as the streaming instability and
hence planetesimal formation.
\end{abstract}


\section{Introduction}

In this paper we examine a phenomenon in which turbulence transports particles down temperature gradients: turbulent thermal
diffusion (TTD), which was originally recognized by \cite{1996PhRvL..76..224E} and subsequently verified in laboratory experiments
\citep{2004NPGeo..11..343E,2006ExFl...40..744E}.
The TTD has astrophysical consequences. For example, planetary atmospheres have vertical varying temperature profiles which play a major
role in the formation of hazes and clouds via condensation \citep{2001ApJ...556..872A}.
The TTD is known to trap particles in such temperature bands in the Earth's atmosphere
\citep{2009JGRD..11418209S}; and has been hypothesized to do
so elsewhere in the Solar System \citep{1997P&SS...45..923E}.

Protoplanetary discs also have temperatures that vary strongly on both global and local scales,
as has been seen in observations of protoplanetary disks, and as has been demonstrated by
laboratory examinations of meteorites \citep{1990Metic..25..309H,2005ASPC..341..353D}.
While models of the background disk temperature, including both heating by irradiation from the central star and
heating from accretion power, predict temperatures that scale as the square-root of the orbital position,
numerical simulations of turbulent accretion disks
that focused on the gas's thermal behavior have seen order unity temperature fluctuations on far smaller, scale-height
length scales \citep{2014ApJ...791...62M}.
Indeed, large local temperature gradients have many possible sources, including
shadowing \citep{2000A&A...361L..17D}, and transitional regions
such as the boundary between magnetically active and magnetically dead regions \citep{2009ApJ...701..620Z},
evaporation fronts \citep{2010ARA&A..48..205D}, or the edges of planet-opened gaps \citep{2012ApJ...748...92T}.
A natural question that arises is to what extent these global and local temperature variations
influence the nature and dynamics of solids in such disks, and thereby also influence planet formation. In particular,
the TTD could play a significant role in the transport and concentration of particles in discs, similarly to how it
acts in planetary atmospheres.

The smaller  spatial scale temperature fluctuations that occurred in our own Solar Nebula,
 and presumably occur in other protoplanetary disks, have consequences beyond the \TTD, i.e.~the
 thermal processing of solid material, \citep[e.g.~chondrule formation,][]{1997AREPS..25...61H}.
 Sufficient heating will cause fluffy particles to compactify, reducing its effective surface area and thereby reducing (although not eliminating)
 its interaction with the gas. This accelerates the rate with which the compactified particles settle or drift radially \citep{2014Icar..237...84H}.
 In addition, temperature fluctuations can also move particles more directly such as through photophoresis,
 which has been invoked to explain Solar Nebula phenomena such as the outwards radial transport of CAIs
 \citep{1918AnP...361...81E,2009M&PS...44..689W}. In an optically
 thick disk, a particle radiatively exchanges thermal information with a shell an optical depth in radius.
 Different sides of a particle can therefore see gas at different temperatures,
 which sets up temperature gradients through the particle. The difference in temperature across the particle's surface cause
 gas molecules to recoil more violently from the hotter side, pushing the particles from hot regions to cold, with
 consequences for protoplanetary disks \citep{2015arXiv151003427M}. However, photophoresis depends on the radiative flux,
 as so requires a high background temperature in addition to a strong temperature gradient.

In this paper we focus only on the \TTD, where, as the name suggests, it is turbulence rather than the radiation field
that plays the central role. Our purpose is two-fold:
firstly, we introduce the TTD to the astrophysics
community and develop it in a more familiar language. In doing so we write the conditions for the TTD to apply in terms sufficiently
general that they can be easily adapted to systems where the transport of solids by gas are important,
including exoplanetary and brown dwarf atmospheres.
Secondly, we explore TTD's consequences
for particle transport in protoplanetary disks in particular and show that 
while background power-law temperature gradients are too weak to support the TTD,
cold annuli about a local scale height wide
are expected to significantly concentrate millimeter to centimeter scale particles in protoplanetary disks.

\section{Turbulent Thermal Diffusion}
\label{Sec_TTD}

\begin{figure}\begin{center}
\includegraphics[width=\columnwidth]{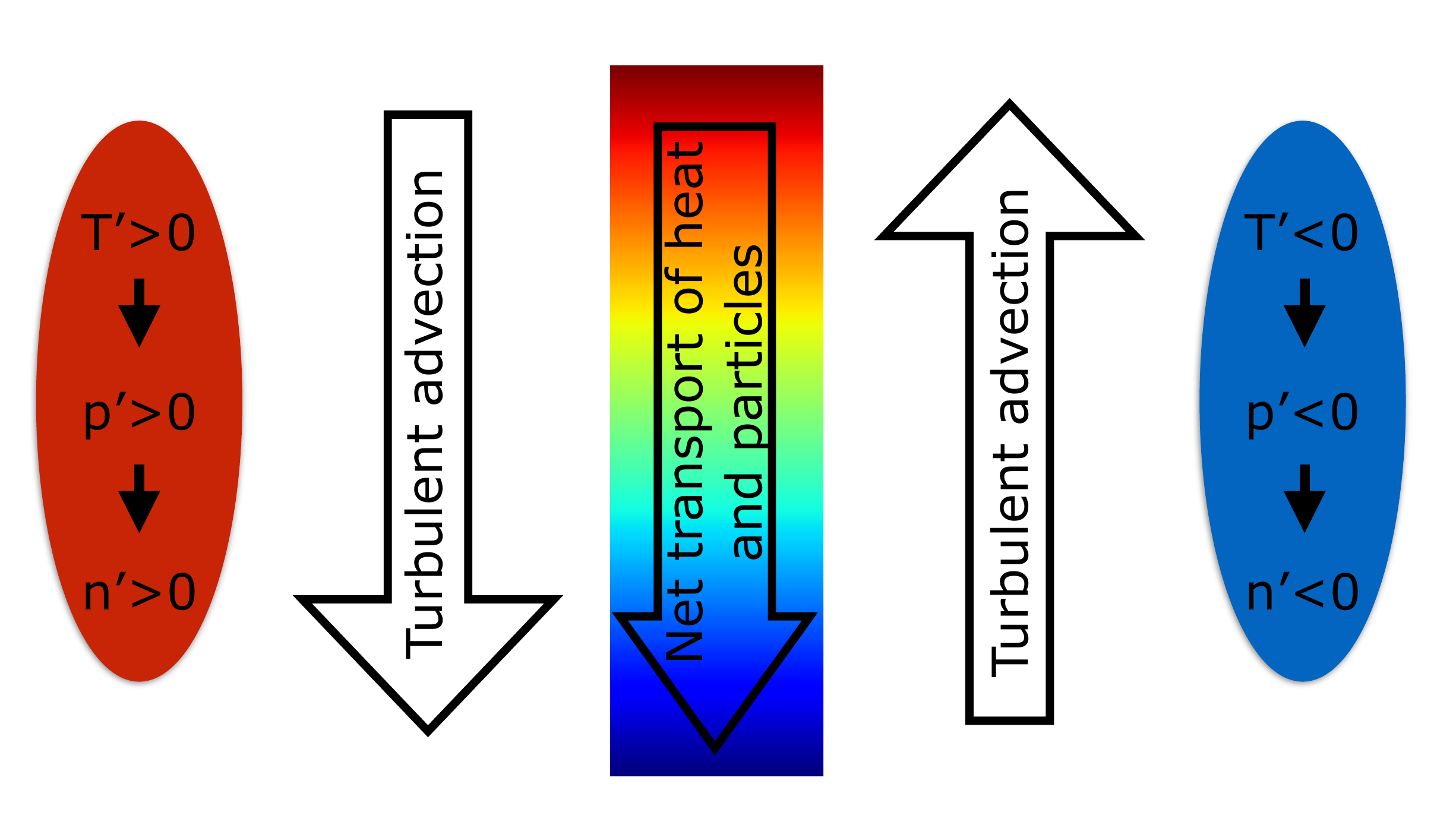}
\end{center}\caption{
Cartoon of the TTD. Turbulent motions moving down (up) a background temperature gradient drags
gas to colder (warmer) regions. The warmer (colder) turbulently advected gas has higher (lower) pressure
than the ambient gas. The higher (lower) pressure turbulently advected gas concentrates (disperses) particles,
resulting in net particle transport.
\label{cartoon} }
\end{figure}

As we will quantify, in general, the TTD acts as a first order effect when all of the following are satisfied:
(1) the temperature gradient is steeper than the pressure gradient, (2) particles are well, but not perfectly, coupled to the turbulence,
and (3) the turbulence is subsonic. The latter two conditions are linked: the better particles are coupled to the turbulence, the more
subsonic the turbulence needs to be. Finally, while the TTD can act on very well coupled particles if the turbulence is sufficiently weak,
the time scale for it to do so can be prohibitive.

Inertial particles, with finite mass $m_p \gg m$, the mean molecular mass of the gas they are embedded in, drift through the gas even if they are frictionally well coupled.
We write the equations for $\ww$, the particle's drift speed through the gas (Equations~\ref{ww_def} and \ref{par_t_w}), as
\begin{align}
&\ww \equiv \vv-\uu, \\
&\partial_t \ww = -\frac{\ww}{\tau_s} + \cdots,
\end{align}
where $\vv$ is the particle's velocity, $\uu$ the gas velocity at the particle's position, and $\tau_s$ the particle's frictional stopping time.

Because even well coupled particles embedded in the gas are too large to feel pressure forces,
they drift through the gas up pressure gradients according to
\EQ
\ww = \frac{\tau_s}{\rho} \nab p, \label{ww_blah}
\EN
where $\rho$ is the gas density.
Taking the divergence of \Eq{ww_blah}, we can see that pressure maxima concentrate particles, and indeed
\EQ
n' \simeq t_p \partial_t n' \simeq t_p \nab \cdot n\ww \simeq   n \frac{t_p \tau_s}{\rho} \nab^2 p \propto \frac{n \tau_s t_p}{\rho} \frac{p'}{l_p^2},
\label{part_t_n_scaling}
\EN
where $n$ is the background particle number density, $n'$ the fluctuating particle number density,
$p'$ a pressure fluctuation,  $l_p$ its length scale and $t_p$ its life time.
In the presence of turbulence, there is a
fluctuating velocity field $\uu'$ which creates a fluctuating pressure field $p'$ which in turn generates a fluctuating particle
number density field $n'$.  This $n'$ is the signature of preferential concentration \citep{1987JFM...174..441M,2001ApJ...546..496C};
but in the absence of large scale gas gradients, symmetry 
means that there is no preferred direction for averaged vector quantities to be aligned with.
As a result,
\EQ
\langle n' \uu' \rangle \propto \langle  p' \uu' \rangle =0,
\EN
and there is no large scale particle transport.

The presence of a large scale gas temperature gradient strongly alters the situation. The key insight of \cite{1996PhRvL..76..224E} was that for
gas in hydrostatic equilibrium, the fluctuating gas density and velocity fields are at most weakly correlated because there is
no net turbulent transport of gas mass even in the presence of a gas density gradient; but in the presence of a gas temperature gradient,
there is net turbulent transport of heat down the temperature gradient. Therefore, the correlation of $\uu'$ and $p'$ can be approximated as
(Equation~\ref{p'}):
\begin{align}
\langle p' \uu' \rangle &= \left\langle  \frac{k_B}{m}\left(\rho T' + \rho' T\right) \uu' \right\rangle \nonumber \\
&\simeq \left\langle  \frac{\rho k_B T'}{m}  \uu' \right\rangle \simeq \left\langle  \frac{\rho k_B (-t_t \uu' \cdot \nab T)}{m}  \uu' \right\rangle \neq 0,
\label{p'u'}
\end{align}
where $t_t$ is the turbulent correlation timescale.
Invoking \Eq{part_t_n_scaling} we find our expected scaling:
\EQ
\langle n' \uu' \rangle \propto n \frac{\tau_s t_p}{l_p^2} \langle p' \uu' \rangle \propto -n \frac{\tau_s k_B}{m} \nab T, \label{scalings}
\EN
where we have used $t_p = t_t$, and $|\uu'| \simeq l_p/t_p$ for turbulent fluctuations.
Accordingly, \Eq{p'u'} implies that turbulence pumps particles from hot regions
to cold ones, as sketched in \Fig{cartoon}.
In essence, the TTD acts by having turbulent, small scale pressure gradients generate local clumps of particles, and
then having global scale temperature gradients lead to those clumps moving in an ordered fashion.

The temperature fluctuations $T'$ invoked above are not due to local sonic compression but rather to
turbulence advecting gas of different temperatures ($T' \simeq -t_t \uu' \cdot \nab T$ for an appropriate
turbulent correlation
time $t_t$). 
This leads to a key aspect of the TTD: $\langle p' \uu'\rangle \propto \langle \uu'^2\rangle$, so $p'$ depends on the turbulent
speed linearly rather than quadratically as is generally the case (e.g. dynamic pressure and Bernoulli's principle).
Because presence of a gas temperature gradient strongly enhances the magnitude of $p'$, it also leads
to enhanced preferential concentration beyond the scope of this paper \citep{2010PhRvE..81e6313E}.
While we do not known of any simple calculation showing why the gas avoids relaxation to more modest
pressure fluctuations,
in \Sec{sec_small_p'} we further explore the consequences of assuming
\EQ
\frac{p'}{p} \propto \Ma^2, \label{p'_quad}
\EN
where $\Ma$ is the turbulent Mach number. We find that, in that regime,
only minor particle transport occurs. Laboratory experiment and atmospheric observations have found
strong effects, implying that the approximations in \Eq{p'u'} are indeed appropriate and that \Eq{p'_quad}
should not be used
\citep{2004NPGeo..11..343E,2009JGRD..11418209S}.

\subsection{Equations of the TTD}
\label{TTD_Eqs}

We derive the TTD beyond the scaling of \Eq{scalings} in Appendix \ref{TTD}. The derivation is
highly involved (note the equation numbers referenced), and is not needed to explore its astrophysical consequences,
so we limit ourselves to using those results.
We therefore quote Equations~\ref{par_t_nbar_8}, \ref{Fn_1b}, \ref{VTTD_3} and \ref{alpha_final}
to encapsulate the TTD (up to approximations made explicit in Appendix \ref{TTD}):
\begin{align}
&\partial_t n + \nab \cdot \FF_n  \simeq 0, \label{dust_continuity}\\
&\FF_n =n \left( \DDD \nab \left[\ln \rho-\ln n\right] + \ww +\VV_{\TTD} \right), \label{Fn_general} \\
&\VV_{\TTD} =- \frac 43 C \tau_s \frac{ k_B T}{m} \ln \St^{-1}  \nab \ln T  \label{VTTD_general}, \\
&\ww = \frac{\tau_s}{\rho} \nab p, \label{wbar_general} 
\end{align}
where $n$ and $\FF_N$ are the particle number densities and fluxes (not normalized to the gas density);
$\tau_s$ and $\St$ are the particle stopping time and Stokes number normalized to the integral scale of the turbulence (Equation~\ref{St_def});
$\rho$, $p$, $T$ and $m$ are the gas density, pressure, temperature and mean molecular mass;
$\DDD$ is the turbulent diffusion coefficient; 
and $C$ is a coefficient of order unity. The main difference between \Eq{scalings} and \Eq{VTTD_general}, the factor
of $\ln \St^{-1}$, is due to the fact that turbulence has a power spectrum, rather than specific values for its velocity, length scale
and time scale.

Note that our analysis has been performed in
the $\St \ll 1$ limit (i.e. particle frictional stopping times much shorter than turbulent correlation times)
to allow scale separation between the dust stopping time $\tau_s$ and the turbulent correlation time scales.
We have also assumed an approximately adiabatic equation of state, with thermal relaxation negligible
on turbulent advective time scales. Thermal relaxation on timescales comparable with the turbulent advective
time scale would add a prefactor below unity to \Eq{T'}, reducing our estimate for turbulent thermal fluctuations; while thermal
relaxation on time scale far shorter than the turbulent time scale would eliminate the effect.

$\VV_{\TTD}$, the particle turbulent thermal diffusion velocity, is the focus of
this paper.  Note that both $\ww$ and $\VV_{\TTD}$ are proportional to $\tau_s$, but
$\ww \neq 0 $ requires the existence of a large scale pressure gradient, while $\VV_{\TTD} \neq 0$ relies
on the existence of a large scale temperature gradient. For an ideal gas, those gradients often are, but need not be,
related; and even when both exist, they need not be aligned or anti-aligned,
so the two velocities can act in concert, or in opposition.
Note also that the form for $\VV_{\TTD}$
in \Eq{VTTD_general} does not make reference to the turbulent velocity scales, which have cancelled out. As discussed in 
\Sec{vel_limit}, our analysis is performed in the limit $n'/n \ll 1$, which implies that $V_{\TTD} \ll u_0$, the turbulent velocity
at the integral scale. When that constraint is violated \Eq{VTTD_general} cannot be used,
but the fluctuating particle number density field $n'/n$ is not small and must therefore nonetheless be treated with care.

In general we can have both large scale temperature and pressure gradients, and so
need to consider both $\VV_{\TTD}$ and $\ww$. 
For the TTD to be a first order effect, we need
$|\VV_{\TTD}| \gtrsim |\ww|$. We therefore determine the conditions for the TTD to be important in 
\Sec{subsecgen} and derive the TTD in the special case of $\ww \sim \boldmath{0}$ in \Sec{subseciso}.

\subsection{General case}
\label{subsecgen}

From \Eqs{VTTD_general}{wbar_general}
we have
\begin{align}
&\VV_{\TTD} = -\frac 43 \frac{C \tau_s k_B T}{m} \ln \St^{-1} \frac{\ee_T}{L_T} \label{VTTD_noniso}, \\
&\ww = \frac{\tau_s}{\rho} p \frac{\ee_p}{L_p} = \frac{\tau_s k_B T}{m} \frac{\ee_p}{L_p} , \label{ww_noniso}
\end{align}
where $\ee_p$ and $\ee_T$ are the directions of the pressure and temperature gradients while
$L_p$ and $L_T$ are the lengthscales over which the pressure and temperature vary:
\begin{align}
&L_p = \left|\nab \ln p \right|^{-1}, \label{Lp} \\
&L_T = \left|\nab \ln T \right|^{-1}. \label{LT}
\end{align}
We can write two alternative conditions for the TTD to be an important player:
\begin{align}
&\left|\VV_{\TTD}\right|\gtrsim \left|\ww\right|, \label{VTTD_noniso_condition1} \\
&\left|\VV_{\TTD}^2\right|\gtrsim \left|\VV_{\TTD} \cdot \ww\right|, \label{VTTD_noniso_condition2}
\end{align}
where the first condition only compares the magnitudes of the two velocities and the second
also considers the degree of alignment.

Using
\Eqs{VTTD_noniso}{ww_noniso}, \Eqs{VTTD_noniso_condition1}{VTTD_noniso_condition2} become:
\begin{align}
&\frac 43 \frac {L_p}{L_T}C \ln \St^{-1} \gtrsim 1, \label{VTTD_noniso_cond}\\
&\frac 43 \frac {L_p}{L_T}C \ln \St^{-1} \gtrsim  \left| \ee_p \cdot \ee_T\right|.
\end{align}
In what follows we will consider only the more restrictive \Eq{VTTD_noniso_cond}.

\subsection{Isobaric case}
\label{subseciso}

If \Eq{VTTD_noniso_cond} is very well satisfied, then we can neglect the particle drift velocity $\ww$ and note that $L_p \gg L_T$,
(i.e.~the system is effectively isobaric).
This allows us to simplify
\EQ
\nab \ln \rho = \nab \ln p - \nab \ln T \simeq - \nab \ln T.
\EN
Using that simplification, \Eqs{Fn_general}{VTTD_general} become
\begin{align}
&\FF_n =-n \DDD \left(  \nab \ln n +\alpha \nab \ln T\right), \label{Fn_isobaric}\\
&\alpha \equiv 1+\frac{4C}{3 \DDD} \tau_s \frac{ k_B T}{m} \ln \St^{-1}, \label{alpha_definition} \\
&V_{\TTD} =-(\alpha-1) \DDD \nab \ln T, \label{VTTD_isobaric}
\end{align}
where $(\alpha-1)$ measures the relative strength of turbulent diffusion and turbulent thermal diffusion. We use the definition
in \Eq{alpha_definition} to follow existing literature \citep[e.g.][]{2004NPGeo..11..343E}; and note that it is the natural parameter for experiments where the gas temperature
and the particle number density
are easy to measure directly, but the gas density and particle concentration, are not.

\section{Consequences of Turbulent Thermal Diffusion}

\subsection{Isobaric equilibrium distribution}

As has been previously calculated \citep[e.g.][]{2006ExFl...40..744E}, we can use \Eq{Fn_isobaric} to find that the
steady state particle distribution ($\FF_n=0$):
\EQ
n \propto T^{-\alpha} \propto \rho^{\alpha}. \label{nproptoT-alpha}
\EN
For non-inertial particles perfectly coupled to the gas, $\tau_s=0$, so $\alpha=1$ and $n \propto \rho$:
as expected, turbulent mixing drives the system to a uniform particle concentration.
For inertial particles, $\tau_s > 0$, $\alpha>1$,
\EQ
\frac{n}{\rho} \propto T^{-(\alpha-1)},
\label{conc_equation}
\EN
and colder regions concentrate particles.

Experiments have found values up to about $\alpha \simeq 2.7$
\citep{2006ExFl...40..744E} while observations of aerosols in the troposphere indicate that $\alpha>20$ occurs in nature
\citep{2009JGRD..11418209S}.
Note that our analysis assumes that the dust fluid density is always much less than that of the gas, so that
any drag the dust exerts on the gas can be neglected. Large values of $\alpha$ could imply local concentrations
of dust large enough exceed that limit. Our analysis can no longer be applied once such conditions occur, but those conditions would also
allow other processes to dominate \citep[e.g.][]{2007Natur.448.1022J}. In situations where the background dust fluid density is too low to backreact
on the gas, but the TTD would generate concentrations of dust sufficiently dense to backreact on the gas, the TTD
can be invoked as a trigger for processes such as the streaming instability.

\subsection{Conditions for relevance}

While the parameter $C$ has not yet been determined, and likely varies depending on the nature of the turbulence,
we expect $C<1$ and \cite{2007BoLMe.125..167Z} suggests $C \sim 0.3$. In that case, \Eq{VTTD_noniso_cond} becomes
\EQ
\St \lesssim e^{-2.5 L_T/L_p}. \label{C=0.3case}
\EN
The shorter the temperature length scale is compared to the pressure length scale, the larger the particles
can be and still get transported by the TTD. This is important because the TTD transports larger particles faster,
and larger particles are more likely to engage in collection behavior such as the streaming instability.
In astrophysical cases of interest such as planetary atmospheres in hydrostatic equilibrium or protoplanetary
disk midplanes, temperature, density and pressure generally vary in the same direction and on comparable
length scales in the absence of  specific
phenomena driving more localized fluctuations. From \Eq{C=0.3case} we can see that the limits on $\St$ depend
very strongly on $L_T/L_p$, and so in general, TTD can overcome pressure based particle drift only in the limit of very small $\St$,
or in the presence of local phenomena which force $L_T \ll L_p$ (i.e.~the isobaric limit).

However, the limit of very small $\St$ means particles which are extremely well coupled to the gas
and so generally well mixed by turbulence.
We define the Mach number of the turbulence as
\EQ
\Ma \equiv \frac{u_0^2}{c_s^2} = \frac{u_0^2 \rho}{\gamma p} = \frac{m u_0^2}{\gamma k_B T},
\EN
and use that along with \Eq{D_Kol} to rewrite \Eq{alpha_definition} as
\EQ
\alpha-1 = \frac{4C}{ u_0^2} \frac{\tau_s}{t_0} \frac{ k_B T}{m} \ln \St^{-1} = \frac{8C}{ \gamma} \frac{\St}{\Ma^2}\ln \St^{-1}.
\label{alpha-1}
\EN
If \Eq{C=0.3case} is to be satisfied by decreasing $\St$, then the TTD will have an effect only for weak turbulence ($\Ma^2 < \St$).

TTD therefore generally applies to astrophysical systems with temperature fluctuations on
length scales that are simultaneously much larger than the turbulent length scales, and
much smaller than the length scales associated with pressure ($\ell_0 \ll L_T \ll L_p$). When
the TTD applies, in practice it concentrates and disperses particles with $1 \gg \St > \Ma^2$.

\section{Temperature gradients in protoplanetary disks}

\subsection{Global temperature gradients}

Turbulent thermal diffusion has been invoked to explain aerosol concentrations in the Earth's tropopause, and
could similarly be applied to exoplanetary and brown dwarf atmospheres, including both
vertical and horizontal temperature stratification. Here we explore the TTD's consequences for protoplanetary disk midplanes, 
noting that protoplanetary disks are slightly awkward because anticipated parameters
only marginally satisfy the assumptions built into our analytical analysis.
We assume the scalings of a Hayashi Minimum Mass Solar Nebula \citep[MMSN,][]{1981PThPS..70...35H}.
Note that discs are generally assumed to be statistically symmetric about the midplane, and are azimuthally
periodic, so near the midplane the vertical/azimthal plane is homogenous and can be averaged over, leaving
only the radial variations.

Background disk temperature gradients are shallow power-laws, occurring on orbital length scales, and
as we calculate next, are not expected to support significant TTD.
In a Hayashi MMSN, the midplane temperature and pressure scale with orbital position $R$ as:
\begin{align}
&T \propto R^{-1/2}, \label{TofR}\\
&p \propto R^{-13/4},
\end{align}
from which we find
\begin{align}
&L_T = 2 R, \\
&L_p = \frac{4}{13} R.
\end{align}
Note that in this case, $\ww$ points radially inwards \citep[headwind induced infall,][]{1977MNRAS.180...57W} while $\VV_{\TTD}$ points outwards.
With those values for the global gradients, \Eq{C=0.3case} is satisfied for
\EQ
\St \lesssim  10^{-7},
\EN
i.e.~for particles sufficiently coupled to the gas that no meaningful drift occurs regardless. However, 
for $C=2/3$ we would have $\St \lesssim 7\times 10^{-4}$, which begins to be relevant. Even for $C=0.3$,
the TTD can slow headwind induced infall by a few tens of percents:
if $\St=0.01$ then $\VV_{\TTD} \sim -0.3\, \ww$.

\subsection{Local, quasi-isobaric temperature gradients}

While background protoplanetary disk global temperature gradients are expected to be too weak relative to global pressure gradients for TTD to be
a first order effect, protoplanetary disk temperatures are not expected to follow perfect power-laws, but rather
are expected to also show strong, localized temperature variations on scales much smaller than the orbital length scale.
 For example, if the dominant source of heating
in a disk is irradiation by the central object, then shadowing can occur: when an annulus of the disk heats, it puffs up, shadowing the disk
outside it, which then cools and contracts \citep{2000A&A...361L..17D,2012A&A...539A..20S}. Similarly, turbulence can
generate long lived, quasi-isobaric order-unity temperature,  fluctuations on length scales only a few percent of the orbital position 
\citep{2014ApJ...791...62M}; and boundaries between regions with different chemistry provide highly localized, extremely long lived sharp temperature
gradients \cite{2009ApJ...701..620Z,2010ARA&A..48..205D}.
We can check the requirements
for such local temperature variations to allow the TTD to operate.

For the quasi-isobaric approximations to apply, we need to satisfy \Eq{C=0.3case} for non-negligible values of $\St$.
As noted above, this requires
some form of temperature perturbation on top of the expected background power-law of \Eq{TofR}.
One possible source is turbulent dissipation. Magneto-hydrodynamical turbulence dissipates its energy
in quasi-2D structures known as current sheets; and, in protoplanetary disks, these structures can be very thin, hot,
and in pressure balance with the exterior. \cite{2014ApJ...791...62M} found order unity temperature variation on length
scales only a few percent of the orbital position with negligible associated pressure structures. These structures clearly
would satisfy \Eq{C=0.3case} for particles with large enough $\St$ to slip through the gas on short time scales.

Another possible way to generate temperature annuli is shadowing \citep{2000A&A...361L..17D,2012A&A...539A..20S}.
In this case, we note that the gas midplane density in a vertically isothermal disk is
\EQ
\rho_0 = \frac{\Sigma}{\sqrt{2\pi} H} = \frac{ \Sigma \Omega_K}{\sqrt{2 \pi} c_s}, \label{rho0}
\EN
where $\Sigma$ is the gas surface density, $H$ the pressure scale height and $\Omega_K$ the local Keplerian frequency. We also have
\EQ
c_s^2 = \frac{\gamma k_B T}{m}. \label{csofT}
\EN
It follows that the midplane pressure is given by
\EQ
p_0 = \frac{\rho_0 c_s^2}{\gamma} = \sqrt{\frac{k_B T}{2 \pi \gamma m}} \Sigma \Omega_K. \label{p0}
\EN
While the onset of shadowing will have a major impact on the temperature profile, strong Coriolis forces imply that
shadowing is not expected to significantly redistribute mass radially.
As long as the temperature gradient caused by shadowing obeys $L_T^{-1} \gg \partial_R \ln \Sigma \sim R^{-1}$ we can
therefore use Equations~(\ref{Lp}), (\ref{LT}) and (\ref{p0}) to approximate
\EQ
L_p \sim 2 L_T.
\EN
In that case \Eq{C=0.3case} would require $\St \lesssim 0.24$.  That upper limit is safely above values associated with
the fragmentation or bouncing barriers, and is safely large enough for the streaming instability to act
\citep{2010A&A...513A..57Z,2015A&A...579A..43C}.

\section{Quasi-isobaric protoplanetary test case}
\label{QITC}

The potential range of parameters for protoplanetary disks is too large to fully analyse here.
Instead, we derive a test case that shows that there are plausible regions of parameter space in which
the TTD would significantly concentrate dust in protoplanetary disks. The constraints we use
to choose the test-case parameters can be used to check other models for the important of the TTD.

\subsection{Quasi-isobaric particle concentration}

In the common $\alpha$-disk model, protoplanetary disks are assumed to accrete due to turbulent stresses,
with a corresponding viscosity $\alphaSS c_s H$, where $\alphaSS$ represents the Shakura-Sunyaev $\alpha$ \citep{1973A&A....24..337S},
not to be confused with the $\alpha$ of \Eq{alpha_definition}.
We assume here that the turbulent viscosity $\alphaSS c_s H$ is approximately equal
to the turbulent diffusivity $\DDD$, i.e.~a Schmidt number $\text{Sc} \sim 1$ \citep{2006MNRAS.370L..71J}.
The time scale of the turbulence driven by orbital shear is also generally
assumed to lie around $\Omega_K^{-1}$ \citep{2006A&A...452..751F}.
 It follows that
\EQ
u_0 \sim \sqrt{\alphaSS} c_s, \label{u_alpha_approx}
\EN
and hence
\EQ
\Ma^2 \sim \alphaSS.
\EN
Accordingly, \Eq{alpha-1} becomes
\EQ
\alpha-1 \sim 1.7 \frac{\St}{\alphaSS} \ln \St^{-1}, \label{alpha_approx}
\EN
where we have assumed $C=0.3$ and $\gamma=1.4$. We can see that significant TTD effects are expected only
for $\St>\alphaSS$, and that \cite{2012Icar..220..162J}'s parameter $S \equiv \St/\alphaSS$ which measure dust transport
through gas is very relevant here as well.

Our current understanding of planet formation has difficulties in growing grains with $\St \sim 10^{-3}$. 
Turbulently driven collisions are expected to be at best growth neutral, and often destructive 
for such grains \citep{2010A&A...513A..57Z}.
While the precise size at which collisional growth fails is uncertain, our current picture is that once grains grow large
enough ($\St \sim 10^{-2}$), collective dust-gas instabilities such as the streaming instability \citep[SI,][]{2007Natur.448.1022J}
take hold, collecting dust into gravitationally unstable clumps which collapse, forming planetesimals directly. However, this
leaves us with a clear dust size gap between the end collisional growth and the triggering of the SI. One further difficulty for the SI
is that it requires background dust concentrations that are significantly super-solar, and the degree
of metallicity enhancement over solar needed is strongly $\St$ dependent. Indeed, \cite{2015A&A...579A..43C}
found that the smallest grains which can trigger the SI have $\St \sim 3 \times 10^{-3}$; and that requires a background
super-solar dust concentration of about a factor of 5 outside of the water ice line, and about a factor of 15 inside of it.
The TTD provides a possible route to generating those super-solar dust concentrations.

The bouncing barrier is expected to occur for particles colliding at approximately $0.1 - 1$\,cm\,s$^{-1}$ and the fragmentation barrier
for particles colliding at around $100$\,cm\,s$^{-1}$  \citep{2010A&A...513A..56G}. Turbulently induced collisions
have a range of possible speeds however, so particles with characteristic collision speeds that
lie between the bouncing and fragmentation barriers are expected to grow slowly through rare collisions at the low end of the
speed distribution \citep{2012A&A...544L..16W}. This leads to a pile up in grain size as growth becomes ever less efficient.
We assume particles with $\St =3 \times 10^{-3}$ \citep[the smallest which can trigger the SI,][]{2015A&A...579A..43C}
and $\alphaSS= 10^{-3}$. Disk thermal speeds in regions of terrestrial planet formation
are of order $c_s = 10^5$\,cm\,s, and expected turbulent dust-dust collision speeds are of order \citep{2013MNRAS.432.1274H}:
\EQ
v_{\text{collision}} \sim 0.25 \sqrt{\St \alphaSS} c_s \sim 40 \text{ cm s}^{-1}.
\EN
This is large enough to lead to bouncing, but likely not fragmentation, and so lies in the regime associated with a size pile up.

For $\St=3 \times 10^{-3}$ and $\alphaSS= 10^{-3}$, \Eq{alpha_approx} estimates $\alpha \sim 30.6$, which through \Eq{nproptoT-alpha}
would imply extreme concentrations of particles in cold regions through the TTD. Indeed, a
temperature perturbation with amplitude $f_T T$, over a length $\ell_T$ has a corresponding $L_T= \ell_T/f_T$;
so $f_T=0.2$, $\ell_T = 0.6H$ satisfies $L_T = 3H \simeq 0.1 R \ll R$, and stronger temperature gradients, at quasi-constant pressure, have been
seen in simulations of MHD turbulence in protoplanetary disks \citep{2014ApJ...791...62M}.
With those values, \Eq{conc_equation}
implies a particle concentration
by a factor of
\EQ
T^{\alpha-1} \simeq (1+f_T)^{\alpha-1} \simeq (1+0.2)^{30.6-1}\simeq 221.
\label{conc_factor_1}
\EN
This is easily enough to have strong effects on the behavior of dust in protoplanetary disks,
and so the TTD could act as a trigger for the SI.
However these values of $\St$, $\alphaSS$, $f_T$ and $\ell_T$ run into the limits
of our analysis.

\subsection{Test case limitations}

Note that \Eq{u_alpha_approx} also implies
\EQ
\ell_0 \simeq \sqrt{\alphaSS} H,
\EN
We require $\ell_0 \ll \ell_T$.
In this test case, this constraint becomes
\EQ
\ell_0 = 0.03H \ll \ell_T = 0.6H,
\EN
which is well satisfied, but stronger turbulence (more than an order of magnitude
larger $\alphaSS$) has length scales long enough that  for small values of $f_T$  we can
only marginally linearize the background temperature field, as was done in Appendix~\ref{TTD}.

More problematically, from \Eq{VTTD_isobaric} we have
\EQ
|\VV_{\TTD}| = (\alpha-1) \alphaSS c_s \frac{H}{L_T} > \frac 13 (\alpha-1) \alphaSS c_s.
\label{VVTTDALPHASS}
\EN
For $\St=3 \times10^{-3}$ and $\alphaSS=10^{-3}$, this becomes
\EQ
|\VV_{\TTD}| \sim 0.01 c_s \simeq 0.3 u_0,
\EN
and the requirement $|\VV_{\TTD}| \ll u_0$ is only weakly fulfilled.
This means that \Eq{VTTD_isobaric} is likely moderately overestimating $\VV_{\TTD}$ as discussed in \Sec{vel_limit}.
Because our analysis has $\VV_{\TTD}$ increasing with particle $\St$, $\St \sim 3\times 10^{-3}$ lies near the upper limit for which our analysis applies
when $\alphaSS = 10^{-3}$.

If we are overestimating $\VV_{\TTD}$ by a factor of two, then we still have $|\VV_{\TTD}| > |\ww|$, but the concentration factor (Equation~\ref{conc_factor_1})
drops  from a factor of $221$ to a factor of
\EQ
1.2^{[30.6-1]/2} \simeq 1.2^{14.8} \simeq 15,
\label{conc_factor_2}
\EN
which is on the edge of triggering the SI inside the water ice line.
The timescale for concentration under these conditions is moderate but not negligible. Halving $\VV_{\TTD}$ we find
\EQ
\frac{\ell_T}{0.5 \VV_{\TTD}} \sim 20\, \text{Orbits}.
\EN
However, smaller particles, with lower values of $\St$, will feel the TTD only in the presence of even weaker turbulence,
with a correspondingly larger concentration time scale.

\subsection{Test case thermal relaxation}

A further constraint on this test case is thermal relaxation, in particular any radiative cooling or heating of
turbulent parcels of gas. As noted in \Sec{TTD_Eqs}, if the relaxation time is shorter than the turbulent time, we expect
the TTD to be weakened. As long as we have small temperature fluctuations, and are in the optically thick, dense limit, we can
approximate radiative cooling as a diffusive process. For an MMSN midplane at $R=1$\,AU, we have an approximate
thermal diffusion coefficient \citep{2014ApJ...791...62M}:
\EQ
\mu \simeq 4 \times 10^{12} \left(\frac{T}{270 \text{ K}} \right)^3 \text{cm}^2\text{ s}^{-1}.
\EN
The corresponding thermal relaxation time is
\EQ
\frac{\ell_0^2}{\mu} \simeq 6.25 \times 10^7 \text{s} \left(\frac{T}{270 \text{ K}} \right)^{-3} \simeq 13 \left(\frac{T}{270 \text{ K}} \right)^{-3}  \Omega_K^{-1}.
\label{t_rad_est}
\EN
As long as the temperature is modest, the
 time scale is much longer than the turbulent time scale of $\Omega_K^{-1}$, so the largest scale turbulence is insensitive to radiative
thermal relaxation. Thermal relaxation will effect smaller eddies farther down the turbulent cascade. However, this appears in the equation for $\VV_{\TTD}$
only logarithmically (Equation~\ref{integral_1}), and so is relatively unimportant.

On the other hand, radiative thermal diffusion is a strong function
of temperature, and \Eq{t_rad_est} would seem to limit the TTD to cool regions of the disk. However, the opacity of the gas-dust mixture depends
mostly on the abundance of approximately micron-sized dust grains, and at high temperatures those evaporate, dramatically lowering the opacity
and moving the system into the optically thin limit. Further, in regions with low gas density, the gas and dust temperatures can decouple, making
radiative cooling even less efficient. Those effects would make \Eq{t_rad_est} a significant underestimate for the thermal relaxation time at much higher
temperatures. We therefore expect the TTD  to be effective in cool regions \citep[unlike photophoresis,][]{2015arXiv151003427M}, and possibly
at very high temperatures (e.g.~chondrule formation) in low density regions, but not in warm, high density and opacity regions.

\subsection{Other dust and disc parameters}

As disc parameters diverge from the ones in \Sec{QITC} the situation gets less clear. Larger values of $\alphaSS$ at constant $\St$ both reduce
the strength of the TTD (reducing the ratio $\St/\alphaSS$ in Equation~\ref{alpha_approx}) and increase the length scale
associated with turbulence. This latter would begin to weaken the condition that turbulent length scales be much shorter than
temperature length scales. Weaker values of $\alphaSS$ would reduce the turbulent speed, resulted in 
ever more severe overestimates of $\VV_{\TTD}$ (Equation~\ref{VVTTDALPHASS} and subsequent discussion)
and longer concentration time scales.
Thermal relaxation also becomes more pronounced for smaller values of $\alphaSS$ and their corresponding smaller length scales.

Weaker turbulence than our test case, or larger particles (higher $\St$ values)
 constrains the upper limit on $\VV_{\TTD}$ and the concentration time scale;
 while stronger turbulence threatens the scale separation between
background and turbulent fields. Nonetheless, it is clear that the presence of scale-height-scaled temperature fluctuations in turbulent protoplanetary disks
should generally result in significant concentration of particles with $\St \gtrsim \alphaSS$ on moderate time scales.
Laboratory or numerical studies will be needed to make precise
estimations of the effectiveness of the TTD in protoplanetary disks in practice.

\section{Conclusions}

We have adapted analysis of turbulent thermal diffusion to a form more useful for astronomy. The TTD is a process where the combination of
turbulent and a background pressure gradient act to pump inertial particles from hot regions to cold ones, and it can lead to large
particle concentrations in the latter \citep{1996PhRvL..76..224E}. As such, it acts similarly to the well known concentration of particles in high pressure regions;
with the significant difference that temperature, unlike pressure, is not an inherently dynamical parameter
(although it is usually strongly correlated with dynamical parameters).

While the TTD has already been considered in the context of planetary atmospheres \citep{2009JGRD..11418209S},
we also show that it is expected to act in protoplanetary disks
that have $\sim$ scale height wide radial temperature banding, concentrating $\St \sim 10^{-3}$ particles in the cold bands by factors of tens.
While such a degree of concentration would push our analysis outside of its strict region of applicability,
it would be a sufficient metallicity enhancement to allow the streaming instability to proceed.
This suggests that if cold bands are common
in protoplanetary disks, they would allow the TTD to act to trigger the SI, and hence would be natural regions of planetesimal formation.
Possible sources for such cold annuli include disc self-shadowing and localized
turbulent dissipation \citep{2000A&A...361L..17D,2012A&A...539A..20S,2014ApJ...791...62M}. 
Shadowing and turbulent dissipation are not expected to be stationary for disk evolution time scales, but chemical boundaries,
including ionization and evaporation fronts, could be, and would also drive steep temperature gradients \citep{2009ApJ...701..620Z,2010ARA&A..48..205D}.
This adds to the already significant interest in such regions as possible places for planet formation \citep{2012ApJ...756...62L}.

The effects of TTD can be implemented in numerical simulations and theory as an ad-hoc velocity $\VV_{\TTD}$, but theoretical
estimates are only approximate and factors of a few in $\alpha$ have a large impact
on the concentration of particles (Equations~\ref{nproptoT-alpha}, \ref{conc_factor_1} and \ref{conc_factor_2}).
Implementing TTD into numerical simulations directly will be difficult, and efforts to date have barely been able to detect
the effect \citep{2012PhFl...24g5106H}. Capturing the TTD requires a large dynamical range between the scale of the temperature variation
and the scale of the turbulence, and a further large dynamical range between the integral scale of the turbulence and the scale
of the turbulence which has the same correlation time as the particle's frictional stopping time. Further, strong effects are only expected in low Mach
number turbulence, but depend on gas compressibility, forcing sound waves to be fully captured.

Laboratory experiments can and have constrained $\VV_{\TTD}$ in the Stokes drag regime
appropriate for planetary atmospheres \citep{2004NPGeo..11..343E,2006ExFl...40..744E};
but probing the Epstein drag regime, where particle sizes are smaller than the gas molecular mean free path,
appropriate for protoplanetary disks would require extremely fine particles in a very dilute gas. Nonetheless, this is the most promising
avenue for constraining the numerical factors such as $C$.

\section*{Acknowledgements}
This paper would not have been written without the past efforts of Nathan Kleeorin and Igor Rogachevskii
to explain the turbulent thermal diffusion process.
The research leading to these results received funding from NASA OSS grant NNX14AJ56G.

\bibliographystyle{mnras}
\bibliography{TTD.bib}

\begin{thebibliography}{}
\makeatletter
\relax
\def\mn@urlcharsother{\let\do\@makeother \do\$\do\&\do\#\do\^\do\_\do\%\do\~}
\def\mn@doi{\begingroup\mn@urlcharsother \@ifnextchar [ {\mn@doi@}
  {\mn@doi@[]}}
\def\mn@doi@[#1]#2{\def\@tempa{#1}\ifx\@tempa\@empty \href
  {http://dx.doi.org/#2} {doi:#2}\else \href {http://dx.doi.org/#2} {#1}\fi
  \endgroup}
\def\mn@eprint#1#2{\mn@eprint@#1:#2::\@nil}
\def\mn@eprint@arXiv#1{\href {http://arxiv.org/abs/#1} {{\tt arXiv:#1}}}
\def\mn@eprint@dblp#1{\href {http://dblp.uni-trier.de/rec/bibtex/#1.xml}
  {dblp:#1}}
\def\mn@eprint@#1:#2:#3:#4\@nil{\def\@tempa {#1}\def\@tempb {#2}\def\@tempc
  {#3}\ifx \@tempc \@empty \let \@tempc \@tempb \let \@tempb \@tempa \fi \ifx
  \@tempb \@empty \def\@tempb {arXiv}\fi \@ifundefined
  {mn@eprint@\@tempb}{\@tempb:\@tempc}{\expandafter \expandafter \csname
  mn@eprint@\@tempb\endcsname \expandafter{\@tempc}}}

\bibitem[\protect\citeauthoryear{{Ackerman} \& {Marley}}{{Ackerman} \&
  {Marley}}{2001}]{2001ApJ...556..872A}
{Ackerman} A.~S.,  {Marley} M.~S.,  2001, \mn@doi [\apj] {10.1086/321540},
  \href {http://adsabs.harvard.edu/abs/2001ApJ...556..872A} {556, 872}

\bibitem[\protect\citeauthoryear{{Carrera}, {Johansen}  \& {Davies}}{{Carrera}
  et~al.}{2015}]{2015A&A...579A..43C}
{Carrera} D.,  {Johansen} A.,   {Davies} M.~B.,  2015, \mn@doi [\aap]
  {10.1051/0004-6361/201425120}, \href
  {http://adsabs.harvard.edu/abs/2015A%26A...579A..43C} {579, A43}

\bibitem[\protect\citeauthoryear{{Cuzzi}, {Hogan}, {Paque}  \&
  {Dobrovolskis}}{{Cuzzi} et~al.}{2001}]{2001ApJ...546..496C}
{Cuzzi} J.~N.,  {Hogan} R.~C.,  {Paque} J.~M.,   {Dobrovolskis} A.~R.,  2001,
  \mn@doi [\apj] {10.1086/318233}, \href
  {http://adsabs.harvard.edu/abs/2001ApJ...546..496C} {546, 496}

\bibitem[\protect\citeauthoryear{{D'Alessio}, {Calvet}  \&
  {Woolum}}{{D'Alessio} et~al.}{2005}]{2005ASPC..341..353D}
{D'Alessio} P.,  {Calvet} N.,   {Woolum} D.~S.,  2005, in {Krot} A.~N.,
  {Scott} E.~R.~D.,   {Reipurth} B.,  eds,  Astronomical Society of the Pacific
  Conference Series Vol. 341, Chondrites and the Protoplanetary Disk. p.~353

\bibitem[\protect\citeauthoryear{{Dullemond}}{{Dullemond}}{2000}]{2000A&A...361L..17D}
{Dullemond} C.~P.,  2000, \aap, \href
  {http://adsabs.harvard.edu/abs/2000A%26A...361L..17D} {361, L17}

\bibitem[\protect\citeauthoryear{{Dullemond} \& {Monnier}}{{Dullemond} \&
  {Monnier}}{2010}]{2010ARA&A..48..205D}
{Dullemond} C.~P.,  {Monnier} J.~D.,  2010, \mn@doi [\araa]
  {10.1146/annurev-astro-081309-130932}, \href
  {http://adsabs.harvard.edu/abs/2010ARA%26A..48..205D} {48, 205}

\bibitem[\protect\citeauthoryear{{Ehrenhaft}}{{Ehrenhaft}}{1918}]{1918AnP...361...81E}
{Ehrenhaft} F.,  1918, \mn@doi [Annalen der Physik] {10.1002/andp.19183611002},
  \href {http://adsabs.harvard.edu/abs/1918AnP...361...81E} {361, 81}

\bibitem[\protect\citeauthoryear{{Eidelman}, {Elperin}, {Kleeorin}, {Krein},
  {Rogachevskii}, {Buchholz}  \& {Gr{\"u}nefeld}}{{Eidelman}
  et~al.}{2004}]{2004NPGeo..11..343E}
{Eidelman} A.,  {Elperin} T.,  {Kleeorin} N.,  {Krein} A.,  {Rogachevskii} I.,
  {Buchholz} J.,   {Gr{\"u}nefeld} G.,  2004, Nonlinear Processes in
  Geophysics, \href {http://adsabs.harvard.edu/abs/2004NPGeo..11..343E} {11,
  343}

\bibitem[\protect\citeauthoryear{{Eidelman}, {Elperin}, {Kleeorin},
  {Rogachevskii}  \& {Sapir-Katiraie}}{{Eidelman}
  et~al.}{2006}]{2006ExFl...40..744E}
{Eidelman} A.,  {Elperin} T.,  {Kleeorin} N.,  {Rogachevskii} I.,
  {Sapir-Katiraie} I.,  2006, \mn@doi [Experiments in Fluids]
  {10.1007/s00348-006-0111-3}, \href
  {http://adsabs.harvard.edu/abs/2006ExFl...40..744E} {40, 744}

\bibitem[\protect\citeauthoryear{{Eidelman}, {Elperin}, {Kleeorin}, {Melnik}
  \& {Rogachevskii}}{{Eidelman} et~al.}{2010}]{2010PhRvE..81e6313E}
{Eidelman} A.,  {Elperin} T.,  {Kleeorin} N.,  {Melnik} B.,   {Rogachevskii}
  I.,  2010, \mn@doi [\pre] {10.1103/PhysRevE.81.056313}, \href
  {http://adsabs.harvard.edu/abs/2010PhRvE..81e6313E} {81, 056313}

\bibitem[\protect\citeauthoryear{{Elperin}, {Kleeorin}  \&
  {Rogachevskii}}{{Elperin} et~al.}{1996}]{1996PhRvL..76..224E}
{Elperin} T.,  {Kleeorin} N.,   {Rogachevskii} I.,  1996, \mn@doi [Physical
  Review Letters] {10.1103/PhysRevLett.76.224}, \href
  {http://adsabs.harvard.edu/abs/1996PhRvL..76..224E} {76, 224}

\bibitem[\protect\citeauthoryear{{Elperin}, {Kleeorin}, {Podolak}  \&
  {Rogachevskii}}{{Elperin} et~al.}{1997}]{1997P&SS...45..923E}
{Elperin} T.,  {Kleeorin} N.,  {Podolak} M.,   {Rogachevskii} I.,  1997,
  \mn@doi [\planss] {10.1016/S0032-0633(96)00159-6}, \href
  {http://adsabs.harvard.edu/abs/1997P%26SS...45..923E} {45, 923}

\bibitem[\protect\citeauthoryear{{Fromang} \& {Papaloizou}}{{Fromang} \&
  {Papaloizou}}{2006}]{2006A&A...452..751F}
{Fromang} S.,  {Papaloizou} J.,  2006, \mn@doi [\aap]
  {10.1051/0004-6361:20054612}, \href
  {http://adsabs.harvard.edu/abs/2006A%26A...452..751F} {452, 751}

\bibitem[\protect\citeauthoryear{{G{\"u}ttler}, {Blum}, {Zsom}, {Ormel}  \&
  {Dullemond}}{{G{\"u}ttler} et~al.}{2010}]{2010A&A...513A..56G}
{G{\"u}ttler} C.,  {Blum} J.,  {Zsom} A.,  {Ormel} C.~W.,   {Dullemond} C.~P.,
  2010, \mn@doi [\aap] {10.1051/0004-6361/200912852}, \href
  {http://adsabs.harvard.edu/abs/2010A%26A...513A..56G} {513, A56}

\bibitem[\protect\citeauthoryear{{Haugen}, {Kleeorin}, {Rogachevskii}  \&
  {Brandenburg}}{{Haugen} et~al.}{2012}]{2012PhFl...24g5106H}
{Haugen} N.~E.~L.,  {Kleeorin} N.,  {Rogachevskii} I.,   {Brandenburg} A.,
  2012, \mn@doi [Physics of Fluids] {10.1063/1.4733450}, \href
  {http://adsabs.harvard.edu/abs/2012PhFl...24g5106H} {24, 075106}

\bibitem[\protect\citeauthoryear{{Hayashi}}{{Hayashi}}{1981}]{1981PThPS..70...35H}
{Hayashi} C.,  1981, \mn@doi [Progress of Theoretical Physics Supplement]
  {10.1143/PTPS.70.35}, \href
  {http://adsabs.harvard.edu/abs/1981PThPS..70...35H} {70, 35}

\bibitem[\protect\citeauthoryear{{Hewins}}{{Hewins}}{1997}]{1997AREPS..25...61H}
{Hewins} R.~H.,  1997, \mn@doi [Annual Review of Earth and Planetary Sciences]
  {10.1146/annurev.earth.25.1.61}, \href
  {http://adsabs.harvard.edu/abs/1997AREPS..25...61H} {25, 61}

\bibitem[\protect\citeauthoryear{{Hewins} \& {Radomsky}}{{Hewins} \&
  {Radomsky}}{1990}]{1990Metic..25..309H}
{Hewins} R.~H.,  {Radomsky} P.~M.,  1990, Meteoritics, \href
  {http://adsabs.harvard.edu/abs/1990Metic..25..309H} {25, 309}

\bibitem[\protect\citeauthoryear{{Hubbard}}{{Hubbard}}{2013}]{2013MNRAS.432.1274H}
{Hubbard} A.,  2013, \mn@doi [\mnras] {10.1093/mnras/stt543}, \href
  {http://adsabs.harvard.edu/abs/2013MNRAS.432.1274H} {432, 1274}

\bibitem[\protect\citeauthoryear{{Hubbard} \& {Ebel}}{{Hubbard} \&
  {Ebel}}{2014}]{2014Icar..237...84H}
{Hubbard} A.,  {Ebel} D.~S.,  2014, \mn@doi [\icarus]
  {10.1016/j.icarus.2014.04.015}, \href
  {http://adsabs.harvard.edu/abs/2014Icar..237...84H} {237, 84}

\bibitem[\protect\citeauthoryear{{Jacquet}, {Gounelle}  \& {Fromang}}{{Jacquet}
  et~al.}{2012}]{2012Icar..220..162J}
{Jacquet} E.,  {Gounelle} M.,   {Fromang} S.,  2012, \mn@doi [\icarus]
  {10.1016/j.icarus.2012.04.022}, \href
  {http://adsabs.harvard.edu/abs/2012Icar..220..162J} {220, 162}

\bibitem[\protect\citeauthoryear{{Johansen}, {Klahr}  \& {Mee}}{{Johansen}
  et~al.}{2006}]{2006MNRAS.370L..71J}
{Johansen} A.,  {Klahr} H.,   {Mee} A.~J.,  2006, \mn@doi [\mnras]
  {10.1111/j.1745-3933.2006.00191.x}, \href
  {http://adsabs.harvard.edu/abs/2006MNRAS.370L..71J} {370, L71}

\bibitem[\protect\citeauthoryear{{Johansen}, {Oishi}, {Mac Low}, {Klahr},
  {Henning}  \& {Youdin}}{{Johansen} et~al.}{2007}]{2007Natur.448.1022J}
{Johansen} A.,  {Oishi} J.~S.,  {Mac Low} M.-M.,  {Klahr} H.,  {Henning} T.,
  {Youdin} A.,  2007, \mn@doi [\nat] {10.1038/nature06086}, \href
  {http://adsabs.harvard.edu/abs/2007Natur.448.1022J} {448, 1022}

\bibitem[\protect\citeauthoryear{{Lyra} \& {Mac Low}}{{Lyra} \& {Mac
  Low}}{2012}]{2012ApJ...756...62L}
{Lyra} W.,  {Mac Low} M.-M.,  2012, \mn@doi [\apj]
  {10.1088/0004-637X/756/1/62}, \href
  {http://adsabs.harvard.edu/abs/2012ApJ...756...62L} {756, 62}

\bibitem[\protect\citeauthoryear{{Maxey}}{{Maxey}}{1987}]{1987JFM...174..441M}
{Maxey} M.~R.,  1987, \mn@doi [Journal of Fluid Mechanics]
  {10.1017/S0022112087000193}, \href
  {http://adsabs.harvard.edu/abs/1987JFM...174..441M} {174, 441}

\bibitem[\protect\citeauthoryear{{McNally} \& {Hubbard}}{{McNally} \&
  {Hubbard}}{2015}]{2015arXiv151003427M}
{McNally} C.~P.,  {Hubbard} A.,  2015, preprint, \href
  {http://adsabs.harvard.edu/abs/2015arXiv151003427M} {} (\mn@eprint {arXiv}
  {1510.03427})

\bibitem[\protect\citeauthoryear{{McNally}, {Hubbard}, {Yang}  \& {Mac
  Low}}{{McNally} et~al.}{2014}]{2014ApJ...791...62M}
{McNally} C.~P.,  {Hubbard} A.,  {Yang} C.-C.,   {Mac Low} M.-M.,  2014,
  \mn@doi [\apj] {10.1088/0004-637X/791/1/62}, \href
  {http://adsabs.harvard.edu/abs/2014ApJ...791...62M} {791, 62}

\bibitem[\protect\citeauthoryear{{Shakura} \& {Sunyaev}}{{Shakura} \&
  {Sunyaev}}{1973}]{1973A&A....24..337S}
{Shakura} N.~I.,  {Sunyaev} R.~A.,  1973, \aap, \href
  {http://adsabs.harvard.edu/abs/1973A%26A....24..337S} {24, 337}

\bibitem[\protect\citeauthoryear{{Siebenmorgen} \& {Heymann}}{{Siebenmorgen} \&
  {Heymann}}{2012}]{2012A&A...539A..20S}
{Siebenmorgen} R.,  {Heymann} F.,  2012, \mn@doi [\aap]
  {10.1051/0004-6361/201118493}, \href
  {http://adsabs.harvard.edu/abs/2012A%26A...539A..20S} {539, A20}

\bibitem[\protect\citeauthoryear{{Sofiev}, {Sofieva}, {Elperin}, {Kleeorin},
  {Rogachevskii}  \& {Zilitinkevich}}{{Sofiev}
  et~al.}{2009}]{2009JGRD..11418209S}
{Sofiev} M.,  {Sofieva} V.,  {Elperin} T.,  {Kleeorin} N.,  {Rogachevskii} I.,
   {Zilitinkevich} S.~S.,  2009, \mn@doi [Journal of Geophysical Research
  (Atmospheres)] {10.1029/2009JD011765}, \href
  {http://adsabs.harvard.edu/abs/2009JGRD..11418209S} {114, 18209}

\bibitem[\protect\citeauthoryear{{Turner}, {Choukroun}, {Castillo-Rogez}  \&
  {Bryden}}{{Turner} et~al.}{2012}]{2012ApJ...748...92T}
{Turner} N.~J.,  {Choukroun} M.,  {Castillo-Rogez} J.,   {Bryden} G.,  2012,
  \mn@doi [\apj] {10.1088/0004-637X/748/2/92}, \href
  {http://adsabs.harvard.edu/abs/2012ApJ...748...92T} {748, 92}

\bibitem[\protect\citeauthoryear{{Weidenschilling}}{{Weidenschilling}}{1977}]{1977MNRAS.180...57W}
{Weidenschilling} S.~J.,  1977, \mnras, \href
  {http://adsabs.harvard.edu/abs/1977MNRAS.180...57W} {180, 57}

\bibitem[\protect\citeauthoryear{{Windmark}, {Birnstiel}, {Ormel}  \&
  {Dullemond}}{{Windmark} et~al.}{2012}]{2012A&A...544L..16W}
{Windmark} F.,  {Birnstiel} T.,  {Ormel} C.~W.,   {Dullemond} C.~P.,  2012,
  \mn@doi [\aap] {10.1051/0004-6361/201220004}, \href
  {http://adsabs.harvard.edu/abs/2012A%26A...544L..16W} {544, L16}

\bibitem[\protect\citeauthoryear{{Wurm} \& {Haack}}{{Wurm} \&
  {Haack}}{2009}]{2009M&PS...44..689W}
{Wurm} G.,  {Haack} H.,  2009, \mn@doi [Meteoritics and Planetary Science]
  {10.1111/j.1945-5100.2009.tb00763.x}, \href
  {http://adsabs.harvard.edu/abs/2009M%26PS...44..689W} {44, 689}

\bibitem[\protect\citeauthoryear{{Zhu}, {Hartmann}, {Gammie}  \&
  {McKinney}}{{Zhu} et~al.}{2009}]{2009ApJ...701..620Z}
{Zhu} Z.,  {Hartmann} L.,  {Gammie} C.,   {McKinney} J.~C.,  2009, \mn@doi
  [\apj] {10.1088/0004-637X/701/1/620}, \href
  {http://adsabs.harvard.edu/abs/2009ApJ...701..620Z} {701, 620}

\bibitem[\protect\citeauthoryear{{Zilitinkevich}, {Elperin}, {Kleeorin}  \&
  {Rogachevskii}}{{Zilitinkevich} et~al.}{2007}]{2007BoLMe.125..167Z}
{Zilitinkevich} S.~S.,  {Elperin} T.,  {Kleeorin} N.,   {Rogachevskii} I.,
  2007, \mn@doi [Boundary-Layer Meteorology] {10.1007/s10546-007-9189-2}, \href
  {http://adsabs.harvard.edu/abs/2007BoLMe.125..167Z} {125, 167}

\bibitem[\protect\citeauthoryear{{Zsom}, {Ormel}, {G{\"u}ttler}, {Blum}  \&
  {Dullemond}}{{Zsom} et~al.}{2010}]{2010A&A...513A..57Z}
{Zsom} A.,  {Ormel} C.~W.,  {G{\"u}ttler} C.,  {Blum} J.,   {Dullemond} C.~P.,
  2010, \mn@doi [\aap] {10.1051/0004-6361/200912976}, \href
  {http://adsabs.harvard.edu/abs/2010A%26A...513A..57Z} {513, A57}

\makeatother
\end{thebibliography}

\appendix
\section{Derivation of Turbulent thermal diffusion}
\label{TTD}

\subsection{Equations for Gas and Solids}

We here derive in exhaustive detail the equations for turbulent particle transport equations in general,
and turbulent thermal diffusion more specifically. We start with the continuity and velocity equations for the gas and particle fluids.
The gas density and velocities are denoted $\rho$ and $\uu$ while the particle number density
and velocity are denoted $n$ and $\vv$. In this paper we assume that the particle fluid has a well
defined, single-valued differentiable velocity field. This single-valued velocity approximation
breaks down at small scales, which is
what allows particle-particle collisions to occur, but the deviation from a well defined velocity field is small.

We assume Kolmogorov turbulence with largest (integral) length and velocity scales $\ell_0$ and $u_0$. The
length scale $\ell_0$ is assumed to be much smaller than the length scale associated with
global variations in parameters of relevance (such as density), so that those fields can be linearized. 
We also define the wavenumber $k_0 \equiv \ell_0^{-1}$ and
the time scale $t_0 \equiv \ell_0/u_0$; and assume that $t_0$ is much shorter than global system
evolution time scales, so that changes in the background fields can be neglected on turbulent time scales.

For simplicity we assume that all global quantities vary along the same direction in a linearizable fashion, and
we adopt a local cylindrical coordinate system with the $z$-axis aligned with that direction. We also assume that
the $xy$ plane can be meaningfully averaged over (i.e.~is periodic, closed, or sufficiently large in extent that
boundary terms can be neglected on the time scale of the \TTD). In a planetary atmosphere, the general
case would have $z$ aligned with altitude, with the temperature varying with height on length scales far
shorter than the local pressure scale height, and also far shorter than those associated with latitude or longitude.
In a protoplanetary disk, we would generally place ourselves at the midplane, with $z$ aligned with the radial direction,
noting that the system is periodic in
azimuth and symmetric about the midplane. In this case for us to be able to average over the radial-azimuthal plane
we need the turbulent length scale to be short compared not
only to the radial gradients, but also compared to the local vertical pressure scale height.

\subsection{Particle Drift Velocity}

Our equations are
\begin{align}
\partial_t & \rho + \nab \cdot \left(\rho \uu\right) = 0, \phantom{\frac{1}{\rho}} \label{par_t_rho} \\
\partial_t & n + \nab \cdot \left(n \vv \right) = 0, \phantom{\frac{1}{\rho}} \label{par_t_n}\\
\partial_t & \uu + \uu \cdot \nab \, \uu =  -\frac{1}{\rho} \nab p + \GG,  \label{par_t_u} \\
\partial_t & \vv + \vv \cdot \nab \, \vv = -\frac{\vv - \uu}{\tau_s}  + \GG, \label{par_t_v}
\end{align}
where $p$ is the gas pressure, $\GG$ the acceleration due to gravity (and any other
accelerations which effect the gas and particles equivalently) and $\tau_s$ is the frictional
stopping time for the particles. In this analysis we neglect
the possibility of any other forces which act on the gas and particles differently, and we neglect the back
reaction of the particle drag on the gas. This last requires that the particle fluid mass density be much less than
the gas density, i.e.:
\EQ
m_p n \ll \rho,
\EN
where $m_p$ is the mass of a particle.

Defining the drift of the particle fluid through the gas as
\EQ
\ww \equiv \vv -\uu \label{ww_def}
\EN
we can combine Equations~\ref{par_t_u} and \ref{par_t_v} to write
\EQ
\partial_t \ww + \ww \cdot \nab \left(\ww + \uu\right)  + \uu \cdot \nab \ww = -\frac{\ww}{\tau_s} + \frac{1}{\rho} \nab p.
\label{par_t_w}
\EN
In the case of $\tau_s \ll t_0$, the particles are well coupled to at least the largest scale turbulence. In that case the
particles will rapidly reach their terminal velocity and we can approximate both
\begin{align}
&\partial_t \ww \simeq 0, \\
&\ww \propto \tau_s, \label{w_OO_tau_s}
\end{align}
so that \Eq{par_t_w} becomes
\begin{align}
\ww &\simeq \tau_s \left[ \frac{1}{\rho} \nab p - \ww \cdot \nab \left(\ww + \uu\right)  - \uu \cdot \nab \ww\right], \\
&=  \frac{\tau_s}{\rho} \nab p
+\left[\text{terms in }\tau_s^2 \text{ and higher}\right].
\label{w_0}
\end{align}
Note that turbulence involves motions on a range of size and time scales, and \Eqs{w_OO_tau_s}{w_0}
neglect the effect of
turbulent motions with wave-numbers $k$ high enough that their corresponding time scales $t(k) \lesssim \tau_s$. This is reasonable
because turbulent structures on those size scale do not live long enough to have a significant
effect on the particle trajectories, but this places limits on which wavenumbers can be integrated over as will be noted in \Sec{part_fluxes}.

\subsection{Mean-field decomposition}

As mentioned above, for simplicity we assume that all large scale spatial variations in the background fields are aligned along the $r$-axis.
We perform a mean-field decomposition, with $xy$ averages denoted by overbars
and fluctuations denoted by primes:
\begin{align}
&\ov{\rho}  \equiv A^{-1} \int_{xy}dxdy\,  \rho, \label{av_Cart}\\ 
&\rho'  \equiv \rho-\ov{\rho},
\end{align}
where
\EQ
A \equiv \int_{xy} dxdy.
\EN
Note that the average of a fluctuating quantity is zero:
\EQ
\ov{\rho'} = \ov{\left[\rho-\ov{\rho}\right]}  = \ov{\rho}- \ov{\rho}=0.
\EN
Using this averaging scheme, only $t$ and $z$ derivatives of averaged quantities survive. Note that this averaging
scheme obeys the Reynolds averaging rules, so derivatives commute with averaging.

In the case of a protoplanetary disk annulus at the midplane with the temperature gradient pointing radially, we can adopt a spherical coordinate
system with the pole ($\theta=0$) aligned with the rotation axis of the disk. In that case, at an orbital position $R$, \Eq{av_Cart}
becomes
\EQ
\ov{\rho} \equiv \frac{1}{2 \pi R \times 2 \Delta \theta R} \int_{\theta=\pi/2 -\Delta \theta}^{\pi/2+ \Delta \theta} R d\theta \int_{0}^{2\pi} R \sin \theta d\phi \rho.
\label{av_Sph}
\EN
In \Eq{av_Sph}, $\phi$ is the azimuthal angle and we require simultaneously that vertical extent $R\Delta \theta$ be large enough
compared to turbulent length scales to average over, and small enough compared to $R$ that curvature can be neglected. In 
protoplanetary disks we
expect turbulent lengthscales to be much smaller than the local scale height, which in turn is expected to be much smaller
than the orbital radius, so those conditions can be met.

By horizontally averaging \Eq{par_t_rho} we find
\EQ
\partial_t \ov{\rho} = -\partial_z \left(  \ov{\rho} \ov{\uu} + \ov{\rho' \uu'}\right),
\EN
noting that the net gas mass flux is
\EQ
A^{-1}\int_{xy} dxdy\, \rho \uu = \ov{\rho}\, \ov{\uu} + \ov{\rho' \uu'}.
\EN
A gas quasi-steady-state with $\partial_t \ov{\rho} \simeq 0$ and no net mass flux in $z$ then satisfies
\EQ
\ov{\rho}\, \ov{\uu} + \ov{\rho' \uu'} = 0,
\EN
which means that
\EQ
\ov{\uu} = -\frac{\ov{\rho' \uu'}}{\ov{\rho}} \neq 0 \label{ubar}
\EN
under this averaging scheme; but also that $\ov{\uu}$ is at most second order in the fluctuations.

\subsection{Gas and particle continuity equations}

Splitting \Eqs{par_t_rho}{par_t_n} into their mean and fluctuating components using \Eq{ww_def} we find
\begin{align}
\partial_t \ov{\rho} &+ \nab \cdot \left(\ov{\rho}\, \ov{\uu}\right) + \nab \cdot \left(\ov{\rho'\uu'}\right) =0, \label{par_t_rhobar} \\
\partial_t \rho'        &+ \nab \cdot \left(\ov{\rho} \uu'+\rho'\ov{\uu}\right) + \nab \cdot \left(\rho' \uu'-\ov{\rho'\uu'}\right)=0, \label{par_t_rho'} \\
\partial_t \ov{n}      &+ \nab \cdot \left( \ov{n}\, \left[\ov{\ww}+\ov{\uu}\right] \right)+ \nab \cdot \left( \ov{n'\ww'}+ \ov{n' \uu'}\right)=0, \label{par_t_nbar}\\
\partial_t n'              &+ \nab \cdot \left( \ov{n} [\ww' + \uu']\right) + \nab \cdot \left(n' [\ov{\ww}+\ov{\uu}]\right) \nonumber \\
                                &+\nab \cdot \left(n' \ww'-\ov{n'\ww'}\right) +\nab \cdot \left(n' \uu'-\ov{n'\uu'}\right) =0. \label{par_t_n'}
\end{align}
To proceed we need to use \Eqs{par_t_rho'}{par_t_n'} to estimate $\rho'$ and $n'$ for use in
\Eqs{par_t_rhobar}{par_t_nbar}. We assume that the fluctuations are weak
enough that we only need to track \Eqs{par_t_rho'}{par_t_n'}  to first order in fluctuating quantities, 
noting from \Eq{ubar} that $\ov{\uu}$ is at most second order in the fluctuations. 
We further assume that background gradients are weak enough that $\ov{\ww}$ can be neglected when estimating $n'$.

Under those conditions, we can use the first order smoothing approximation (FOSA) to close our system, writing
\begin{align}
&\rho'(t) \simeq \rho'(0)-\int_0^t dt' \nab \cdot \left[\ov{\rho}(t') \uu'(t') \right], \label{rho'0} \\
&n'(t) \simeq n'(0)-\int_0^t dt' \nab \cdot \left\{ \ov{n}(t') \left[\ww'(t')+\uu'(t')\right]\right\}. \label{n'0}
\end{align}
As long as mean quantities vary slowly compared to turbulent time scales, and noting that the fluctuating quantities vary around
zero, we need track the time
integrals in \Eqs{rho'0}{n'0} only for the turbulent correlation times, finding
\begin{align}
&\rho' \simeq -t_u \nab \cdot \left(\ov{\rho} \uu' \right), \label{rho'} \\
&n' \simeq -t_w \nab \cdot \left( \ov{n} \ww'\right) -t_u \nab \cdot \left( \ov{n}\uu'\right), \label{n'}
\end{align}
for characteristic turbulent time scales $t_w$ and $t_u$. In what follows we will assume that at every turbulent length scale
$t_w = t_u$.

Combining \Eqs{par_t_rhobar}{rho'} and combining \Eqs{par_t_nbar}{n'}, using the fact that mean quantities can only vary in the $z$ direction, we find
\begin{align}
\partial_t \ov{\rho} &+ \partial_z \left(\ov{\rho}\, \ov{u}_z\right)
-\partial_z \left(\ov{t_u u_z' [\ov{\rho} \nab \cdot \uu'+ u_z' \partial_z \ov{\rho} ]} \right)=0, \label{par_t_rhobar2}\\
\partial_t \ov{n} &+ \partial_z \left( \ov{n}\, \ov{v}_z \right) \nonumber \\
&- \partial_z \left(\ov{t_u w_z' [\ov{n} \nab \cdot \ww'+ w_z' \partial_z \ov{n} ] }\right) \nonumber \\
&- \partial_z \left(\ov{t_u w_z' [\ov{n} \nab \cdot \uu'+ u_z' \partial_z \ov{n} ]} \right) \nonumber \\
&- \partial_z \left(\ov{t_u u_z'  [\ov{n} \nab \cdot \ww'+ w_z' \partial_z \ov{n} ]} \right) \nonumber \\
&- \partial_z \left(\ov{t_u u_z'  [\ov{n} \nab \cdot \uu'+ u_z' \partial_z \ov{n} ]} \right)=0. \label{par_t_nbar_2}
\end{align}
As shown in \Eq{ubar}, $\ov{u}$ need not be zero,
especially in the presence of background density gradients. However, $\ov{u}$ can be considered an artifact of choosing an averaging scheme
which obeys the Reynolds averaging rules rather than a density-weighted average, and fortunately, it can be eliminated
from the equations for the dust in favor of turbulent terms:
assuming a gas density statistical steady state, we have $\partial_t \ov{\rho} =0$ and \Eq{par_t_rhobar2} implies that
\begin{equation}
\ov{\rho} \left( \ov{u}_z-\ov{u_z'  t_u \nab \cdot \uu'} \right)= \ov{ t_u u_z'^2} \partial_z \ov{\rho}. \label{stead_state_imp}
\end{equation}
\Eq{stead_state_imp} captures the density evolution of turbulent parcels of gas moving across a background gas density gradient.

Combining and rearranging terms in \Eq{par_t_nbar_2} we arrive at
\begin{align}
\partial_t \ov{n} + \partial_z \left( \ov{n}\, \ov{v}_z \right) 
&-\partial_z \left(\left[\ov{t_u w_z'^2+ t_u u_z'^2+2t_u w_z' u_z'} \right] \partial_z \ov{n}   \right)   \nonumber \\
&- \partial_z \left(\ov{[t_u w_z' +t_u u_z'] \left[\nab \cdot \ww' +\nab \cdot \uu' \right]} \, \ov{n}\right) =0. \label{par_t_nbar_3}
\end{align}

\subsection{Ordering}

We recall from \Eq{ubar}  that $\ov{\uu}$ is at most quadratic in fluctuating quantities. Averaging \Eq{w_0}
we find
\EQ
\ov{\ww} \simeq \frac{\ov{\tau_s}}{\ov{\rho}} \nab \ov{p} + \left[\text{terms quadratic and higher in fluctuations}\right] \label{wbar}.
\EN
Because we have assumed that background quantities vary on long
length scales compared to the turbulence, we can approximate 
\EQ
\ww' \simeq \left(\frac{\tau_s'}{\ov{\rho}}-\frac{\ov{\tau_s} \rho'}{\ov{\rho}^2}\right) \nab \ov{p}
+ \frac{\ov{\tau_s}}{\ov{\rho}} \nab p' \simeq   \frac{\ov{\tau_s}}{\ov{\rho}} \nab p' . \label{w'}
\EN
To help identify terms we define
\begin{align}
&\DDD = \ov{t_u u_z'^2}, \\
&\tilde{\DDD} =\ov{t_u w_z'^2+2t_u w_z' u_z'}, \\
&\tilde{w} = - \ov{t_u w_z' \nab \cdot \uu' }, \\
&\tilde{u} = - \ov{t_u u_z' \nab \cdot \uu' },
\end{align}
where $\DDD$ is the traditional turbulent diffusion coefficient, $\tilde{\DDD}$ a correction for inertial
particles, and $\tilde{w}$ and $\tilde{u}$  are correction terms deriving from our use of a volume (rather than
gas density) weighted averaging scheme.

We can now rewrite \Eq{par_t_nbar_3} as
\begin{align}
\partial_t \ov{n} &+ \partial_z \left( \ov{n}\, [\ov{u}_z+\tilde{u}+\ov{w}_z+\tilde{w}] \right) -\partial_z \left(\DDD \partial_z \ov{n}\right) \nonumber\\
&-\partial_z \left(\tilde{\DDD} \partial_z \ov{n}   \right)  
- \partial_z \left(\ov{t_u [w_z' + u_z'] \nab \cdot \ww' } \, \ov{n}\right) \simeq 0. \label{par_t_nbar_4}
\end{align}
Finally, we can use \Eq{stead_state_imp} to write
\EQ
\ov{u}_z+ \tilde{u} = \ov{t_u u_z'^2} \partial_z \ln \ov{\rho} = \DDD \partial_z \ln \ov{\rho},
\EN
reducing \Eq{par_t_nbar_4} to
\begin{align}
\partial_t \ov{n} &+ \partial_z \left( \ov{n} \DDD\partial_z \left[  \ln \ov{\rho} -  \ln \ov{n}\right] \right) 
+ \partial_z \left( \ov{n}\, [\ov{w}_z+\tilde{w}] \right) \nonumber \\
&-\partial_z \left(\tilde{\DDD} \partial_z \ov{n}   \right) 
- \partial_z \left(\ov{t_u [w_z' + u_z'] \nab \cdot \ww' } \, \ov{n}\right) \simeq 0. \label{par_t_nbar_5}
\end{align}
In \Eq{par_t_nbar_5}, the first two terms are the continuity equation for a passive scalar and the
remaining terms are the corrections for particle inertia. Note that, as expected, in the absence of particle inertia the steady state solution obeys
$\ov{n} \propto \ov{\rho}$, i.e.~a uniform particle concentration.

Turbulence with a wavenumber $k$ has an associated gas velocity $u(k)$. As long as
the turbulent time scale $t(k) \gg \tau_s$, the corresponding relative velocity $w(k) \ll u(k)$,
and we approximate $\ww \ll \uu$. However, the same need not hold when comparing
 $\nab \cdot \ww$  and $\nab \cdot \uu$ because the latter is resisted by pressure forces.
We therefore drop $\tilde{w}$ and $\tilde{\DDD}$ in favor of $\tilde{u}$ and $\DDD$, reducing \Eq{par_t_nbar_5} to
\begin{align}
\partial_t \ov{n} + \partial_z \left( \ov{n} \DDD\partial_z \left[  \ln \ov{\rho} -  \ln \ov{n}\right] \right)
&+ \partial_z \left( \ov{n}\, \ov{w}_z \right)  \nonumber \\
&- \partial_z \left(\ov{t_u u_z' \nab \cdot \ww' } \, \ov{n}\right) \simeq 0. \label{par_t_nbar_6}
\end{align}
Combining \Eqs{w'}{par_t_nbar_6}, repeating the approximation that derivatives on fluctuating quantities dominate over
derivatives on background quantities, we come at long last to
\begin{align}
\partial_t \ov{n} + \partial_z \left( \ov{n} \DDD\partial_z \left[  \ln \ov{\rho} -  \ln \ov{n}\right] \right)
&+ \partial_z \left( \ov{n}\, \ov{w}_z \right) \nonumber \\
& - \partial_z \left(\frac{\ov{\tau_s}}{\ov{\rho}}\ov{t_u u_z' \nabla^2 p' } \, \ov{n}\right) \simeq 0. \label{par_t_nbar_7}
\end{align}
\Eq{par_t_nbar_7} is the general low Mach number turbulent averaged particle fluid continuity equation under standard mean-field decompositions and for
common scale-separation assumptions. We next explore its final term and derive the TTD. 

\subsection{Pressure fluctuations in the presence of a temperature gradient}
\label{part_fluxes}

Using \Eq{par_t_nbar_7} we can define the particle flux $F_n$ in the $\ez$ direction:
\begin{align}
&\partial_t \ov{n} + \partial_z \left(F_n \right) \simeq 0. \label{par_t_nbar_8}\\
&F_n \equiv\ov{n} \left( \DDD \partial_z \left[\ln \ov{\rho}-\ln \ov{n}\right] + \ov{w}_z - \frac{\ov{\tau_s}}{\ov{\rho}} \ov{t_u u_z' \nabla^2 p' }\right).
\label{Fn_1}
\end{align}
In \Eq{Fn_1}, the first term on the right-hand side is the diffusive flux, the second the large scale particle drift due to a background pressure
gradient.  The third term is the source of the so-called turbulent thermal diffusion.

We can use the ideal gas equation of state to write, to first order in fluctuating quantities,
\begin{align}
&\ov{p} = \frac{k_B}{m}\ov{\rho} \ov{T}, \\ 
&p' =  \frac{k_B}{m}\left(\ov{\rho} T' + \rho' \ov{T}\right), \label{p'} \\
& \frac{\ov{\tau_s}}{\ov{\rho}} \ov{t_u u_z' \nabla^2 p' } =  \frac{\ov{\tau_s} k_B}{m \ov{\rho}} \ov{t_u u_z' \nabla^2 \left(\rho' \ov{T}+\ov{\rho} T'\right)} 
\label{pressure_term_approx}
\end{align}
where $m$ is the mean molecular mass of the gas.
The key insight of \cite{1996PhRvL..76..224E} was that the correlation of $u_z'  \nabla^2 \rho'$ in \Eq{pressure_term_approx} is small because it represents turbulent transport of mass,
but the correlation of $u_z' T'$ is large in the presence of a large scale temperature gradient, because it represents the turbulent
transport of temperature down a temperature gradient.

We therefore neglect the $\rho'$ component of \Eq{p'} and write the equation for turbulent thermal diffusion as:
\begin{align}
- \frac{\ov{\tau_s}}{\ov{\rho}} \ov{t_u u_z' \nabla^2 p' } &\simeq  -\frac{\ov{\tau_s}}{\ov{\rho}} \ov{t_u u_z' \nabla^2 \frac{k_B}{m} \ov{\rho} T' } \nonumber \\
&\simeq -\frac{\ov{\tau_s} k_B}{m} \ov{t_u u_z' \nabla^2  T' } \equiv V_{\TTD}, \label{VTTD_1}
\end{align}
where we have neglected $\nabla^2 \ov{\rho}$ because turbulent length scales are assumed shorter than turbulent ones.
In the absence of strong cooling terms, temperature is approximately advected:
\EQ
T' = - C  t_u u_z' \partial_z \ov{T}, \label{T'}
\EN
for a turbulent transport coefficient  $C$ of order unity which depends on the nature of the turbulence \citep{2007BoLMe.125..167Z}. Combining \Eqs{VTTD_1}{T'} we find
\begin{align}
V_{\TTD} &\simeq C\frac{\ov{\tau_s} k_B}{m} \partial_z \ov{T} \, \ov{t_u u_z' \nabla^2  t_u u_z' }, \nonumber \\
&\simeq C\frac{\ov{\tau_s} k_B \ov{T}}{m} \partial_z \ln \ov{T} \, \ov{t_u u_z' \nabla^2  t_u u_z' } \label{VTTD_2},
\end{align}
where we have again used scale separation to neglect derivatives on non-turbulent gradients.

For Kolmogorov turbulence we have
\begin{align}
&t_u =2 t_0 \left(\frac{k}{k_0}\right)^{-2/3}, \\
&|\uu'| = u_0 \left(\frac{k}{k_0}\right)^{-1/3}.
\end{align}
Accordingly, we can calculate $\DDD$ by integrating over the turbulent cascade:
\EQ
\DDD =  \ov{t_u u_z'^2}  = \int_{k_0}^{k_\eta} B \, 2 t_0 \left(\frac{k}{k_0}\right)^{-2/3}  u_0 \left(\frac{k}{k_0}\right)^{-2/3} \frac{dk}{k}.
\EN
For isotropic turbulence Kolmorov turbulence we have $B=2/9$, and the dissipation wavenumber 
\EQ
k_\eta= Re^{3/4} k_0, \label{St_limit_Re}
\EN
where $Re$ is the Reynolds number of the turbulence. For astrophysical turbulence, we generally have $Re \gg 1$ and so
\EQ
\DDD=\frac{t_0 u_0^2}{3}. \label{D_Kol}
\EN

Replacing $\nabla^2 \rightarrow k^2$ we can also calculate 
\begin{align}
\ov{t_u u_z' \nabla^2  t_u u_z' } &=\int_{k_0}^{k_1}  B\left[4 t_0^2 \left(\frac{k}{k_0}\right)^{-4/3}\right] \left[u_0^2 \left(\frac{k}{k_0}\right)^{-2/3}\right]
 k^2 \frac{dk}{k}, \nonumber \\
&= -\frac 89 \int_{k_0}^{k_1} \frac{dk}{k} =- \frac 89 \ln \left(k_1/k_0\right), \label{integral_1}
\end{align}
where $k_1$ is the limiting wave number.  We have assumed that $\tau_s < t_u$, which imposes
\EQ
\frac{k_1}{k_0} < \left(\frac{1}{\St}\right)^{3/2}, \label{St_limit}
\EN
where
\EQ
St \equiv \frac{\tau_s}{2t_0} \label{St_def}
\EN
is the Stokes number defined with respect to the integral scale of the turbulence.
In astrophysical contexts the Reynolds number $Re$ is typically very large, so in practice \Eq{St_limit} is the controlling
factor in setting the upper limit in the integral in \Eq{integral_1}, rather than the limit
\EQ
k_1 = k_\eta =   Re^{3/4} k_0 
\EN
set by the dissipation scale of the turbulence. Further, $V_{\TTD}$ depends linearly on $\St$ so its effects will be negligible
unless $\St$ is large enough that \Eq{St_limit} is indeed the controlling limit.

In much of the existing literature, \Eq{St_limit_Re} is the limit used, and in the cases where \Eq{St_limit} is invoked,
it is done so in the Stokes drag regime. Many cases of astrophysical interest are in the Epstein drag regime so we proceed using the more general
formulation in \Eq{St_limit}, resulting in the equations for the particle flux:
\begin{align}
&F_n =\ov{n} \left( \DDD \partial_z \left[\ln \ov{\rho}-\ln \ov{n}\right] + \ov{w}_z +V_{\TTD} \right).
\label{Fn_1b} \\
&V_{\TTD} =- \frac 43 C \ov{\tau_s} \frac{ k_B \ov{T}}{m} \ln \St^{-1}  \partial_z \ln \ov{T}  \label{VTTD_3}.
\end{align}

\subsection{Velocity limits}
\label{vel_limit}

An interesting feature of \Eq{VTTD_3} is that the turbulent velocity scale has cancelled out, and for an arbitrarily large $\partial_z \ln \ov{T}$, 
we would have $V_{\TTD} > u_0$, which would be absurd. Returning to \Eq{VTTD_2}, note that
\EQ
\frac{n'_{T'}}{\ov{n}} = C\frac{ \ov{\tau_s} k_B \ov{T}}{m} \partial_z \ln \ov{T} t_u \nabla^2 t_u u_z' \label{vel_constr_1}
\EN
is the fractional particle number density fluctuation due to the temperature fluctuations. We have assumed that the turbulent fluctuations are small
enough that they can be treated to first order, so we require $n'_{T'} \ll \ov{n}$, which in turn constrains $V_{\TTD} \ll u_0$.  For Kolmogorov turbulence,
the right hand side of \Eq{vel_constr_1} is proportional to
\EQ
t_u^2 \nabla^2 u_z' \propto k^{1/3},
\EN
and so the requirement that 
\EQ
 \frac{n'_{T'}}{\ov{n}} = C\frac{ \ov{\tau_s} k_B \ov{T}}{m} \partial_z \ln \ov{T} t_u \nabla^2 t_u u_z' \ll 1
 \EN
 is a requirement on both $\partial_z \ln \ov{T}$ and on the upper limit $k_1/k_0$ for the integral in \Eq{integral_1}. Throughout this paper we
 assume that $\partial_z \ln \ov{T}$ is sufficiently small that \Eq{VTTD_3} can be used. When that is not the case, \Eq{VTTD_3} overestimates
 $V_{\TTD}$, but in that case the concentration of particles invoked by TTD is large enough to be non-negligible on its own terms.

\subsection{Quasi-isobaric case}

In the limit where $\partial_z \ov{p} =0$, we have 
\EQ
\partial_z \ln \ov{\rho} = - \partial_z \ln \ov{T},
\EN
and we can write \Eq{Fn_1} as 
\begin{align}
F_n &= \ov{n} \left( -\DDD \partial_z \left[\ln \ov{T}+\ln \ov{n}\right] - \frac 43 C \ov{\tau_s} \frac{ k_B \ov{T}}{m} \ln \St^{-1}  \partial_z \ln \ov{T}\right) \nonumber \\
&=- \DDD \ov{n} \left( \left[1+\frac{4C}{3 \DDD} \ov{\tau_s} \frac{ k_B \ov{T}}{m} \ln \St^{-1} \right] \partial_z \ln \ov{T} + \partial_z \ln \ov{n}\right).
\label{Fn_2}
\end{align}
Defining
\EQ
\alpha \equiv 1+\frac{4C}{3 \DDD} \ov{\tau_s} \frac{ k_B \ov{T}}{m} \ln \St^{-1}, \label{alpha_final}
\EN
we see that the steady state solution is
\EQ
\ov{n} \propto \ov{T}^{-\alpha},
\EN
where $\ov{n}$ and $\ov{T}$ are observables in both laboratory experiments and the Earth's atmosphere, allowing $\alpha$ to be evaluated.
Values of $\alpha>2$ have been found in experiments and $\alpha \gtrsim 20$ in measurements of aerosols in the Earth's tropopause.
Under these definitions,
\EQ
V_{\TTD}= -(\alpha-1) \DDD \partial_z \ln \ov{T}. \label{VTTD_Final}
\EN

\subsection{Falsifying rapid pressure equilibration}
\label{sec_small_p'}

It might seem natural to assume that low Mach number turbulence experiences rapid pressure equilibration, and therefore that $p'$ is set
by the constraint that vertically traveling parcels of gas rapidly expand or contract to maintain pressure equilibration.
However, as we show here, it that were the case then particle transport would be negligible, 
which is incompatible with the results of laboratory investigations of the TTD.
In the isobaric case, rapid equilibration would imply $p'=0$ in
the absence of thermal equilibration, which has been ruled out by the experiments which found $\alpha>1$.

Assuming instead perfect thermal equilibration results in the following approximations for \Eqs{par_t_rho}{par_t_u}:
\begin{align}
& \partial_t \rho \simeq 0 \simeq -u'_z \partial_z \bar{\rho} -\ov{\rho} \nab \cdot \uu', \label{par_t_rho_small_p'}\\
& \partial_t \nab \cdot \uu' \simeq -\frac {1}{\ov{\rho}} \nabla^2 p'. \label{par_t_div_u}
\end{align}
Under the assumption of small turbulent correlation times we can also estimate
\EQ
\nab \cdot \uu' \simeq t_u \partial_t \nab \cdot \uu'. \label{div_u_par_t_div_u}
\EN
Inserting \Eq{par_t_div_u} into \Eq{div_u_par_t_div_u} we find
\EQ
\nab \cdot \uu' \simeq -\frac{t_u}{ \ov{\rho}} \nabla^2 p'. \label{div_u_par_t_div_u2}
\EN
Combining \Eqs{par_t_rho_small_p'}{div_u_par_t_div_u2}, we arrive at
\EQ
\nabla ^2 p' \simeq \frac{u'_z}{t_u} \partial_z \ov{\rho},
\EN
which when inserted into \Eq{Fn_1} would imply
\EQ
V_{\TTD} \simeq -\frac{\ov{\tau_s}}{\ov{\rho}} \ov{t_u u_z' \frac{u'_z}{t_u} \partial_z \ov{\rho}}
=-\ov{\tau_s} \ov{u_z'^2}\frac{\partial_z \ov{\rho}}{\ov{\rho}} =-\frac{\ov{\tau_s} u_0^2}{3} \partial_z \ln \ov{\rho} ,
\EN
and in the case of Kolmogorov turbulence with zero large scale pressure gradients this reduces to
\EQ
V_{\TTD} \simeq - 2 \St \DDD \partial_z \ov{\rho} = 2 \St \DDD \partial_z \ov{T},
\EN
which has the opposite sign as the prediction of turbulent thermal diffusion. Thus this approximation would predict $\alpha-1<0$ which has been ruled out by observation
and experiment.

We could also assume that $p'$ is controlled by the turbulent ram pressure $\ov{\rho} |\uu'|^2$. 
In the absence of large scale gradients in the gas,  symmetry implies that the ram pressure
fluctuations cannot correlate with $u_z'$ in \Eq{Fn_1}. However, in the presence of a temperature gradient we can postulate 
\EQ
|p'| \sim |\ov{\rho} u_z'^2\,  k^{-1} \partial_z \ln \ov{T}|,
\EN
where the $k^{-1}\ln \ov{T}$ term provides the required isotropy breaking for a correlation to exist. In the case of Kolmogorov turbulence this reduces
to
\EQ
|V_{\TTD}| = |4 \St \DDD \partial_z \ov{T}| \ll |\DDD \partial_z \ov{T}|,
\EN
and so would predict
\EQ
|\alpha-1| \ll 1,
\EN
which also ruled out by experiment and observation.

\end{document}